\documentclass[namedreferences]{solarphysics}

\usepackage[hyperref,optionalrh,showbiblabels]{spr-sola-addons} 
\usepackage{graphicx}        
\usepackage[dvipsnames]{xcolor}
\usepackage{color}           
\usepackage{breakurl}        
\usepackage{xcolor}
\usepackage{multirow}

\chardef\us=`\_

\begin{document}

\begin{article}
\begin{opening}

\title{Generation of low-frequency kinetic waves at the footpoints of pre-flare coronal loops}

\author[addressref={aff1},corref,email={e-mail.alexandr.kryshtal@gmail.com}]{\inits{A.}\fnm{Alexandr}~\lnm{Kryshtal}}
\author[addressref=aff1]{\inits{A.}\fnm{Anna}~\lnm{Voitsekhovska}}
\author[addressref=aff1]{\inits{O.}\fnm{Oleg}~\lnm{Cheremnykh}}
\author[addressref=aff2,email={e-mail.I.Ballai@sheffield.ac.uk}]{\inits{I.}\fnm{Istvan}~\lnm{Ballai}}
\author[addressref=aff2,email={e-mail.G.Verth@sheffield.ac.uk}]{\inits{G.}\fnm{Gary}~\lnm{Verth}}
\author[addressref=aff3,email={e-mail.v.fedun@sheffield.ac.uk}]{\inits{V.}\fnm{Viktor}~\lnm{Fedun}}


\address[id=aff1]{Space Research Institute, Kiev, 03680, 187, Ukraine}
\address[id=aff2]{Plasma Dynamics Group, School of Mathematics and Statistics, The University of Sheffield, Hicks Building, Hounsfield Road, Sheffield S3 7RH, United Kingdom}
\address[id=aff3]{Plasma Dynamics Group, Department of Automatic Control and Systems Engineering, The University of Sheffield, Mappin Street, Sheffield, S1 3JD, United Kingdom}

\runningauthor{Kryshtal et al.}
\runningtitle{Generation of low-frequency kinetic waves in pre-flare loops}

\begin{abstract}
In this study we discuss the excitation of low frequency plasma waves in the lower-middle chromosphere region of loop footpoints for the case when the plasma can be considered in a pre-flare state. It is shown, that among the known semi-empirical models of the solar atmosphere, only the VAL (F) model together with a particular set of basic plasma parameters and amplitudes of the electric and magnetic fields supports generation of low frequency wave instability. Our results show that it is possible to predict the onset of the flare process in the active region by using the interaction of kinetic Alfv\'{e}n and kinetic ion-acoustic waves, which are solutions of the derived dispersion equation. The VAL (F) model allows situations when the main source of the aforementioned instability can be a sub-Dreicer electric field and drift plasma movements due to presence of spatial inhomogeneities. We also show that the generation of kinetic Alfv\'{e}n and kinetic ion-acoustic waves can occur both, in plasma with a purely Coulomb conductivity and in the presence of small-scale Bernstein turbulence. The excitation of the small amplitude kinetic waves due to  development of low threshold instability in plasma with relatively low values of the magnetic field strength is also discussed.
\end{abstract}
\end{opening}

\section{Introduction} \label{sec:intro}

High-resolution solar observational data obtained during Hinode (Solar B), SDO, STEREO, IRIS missions have confirmed that the coronal heating problem will most likely only be solved if the photosphere, chromosphere, transition region and corona are considered together as a connected energetic circuit \citep[see, e.g.][]{Aschwanden2001, Aschwanden2005}. In this regard present theoretical models that attempt to explain the problem of coronal heating have a number of limitations of their applicability. These limitations are related to the spatial inhomogeneity of the plasma \citep[see, e.g.][]{Aschwanden2001, Aschwanden2005, Heinzel2007, Reale2010}, the location of the regions of the primary energy release of some flares in Active Regions (ARs) \citep[][]{Reznikova2009, Zaitsev1994, Charikov2015}, anomalously low temperatures of individual flares \citep[][]{Shchukina2013, Benka1994, Somov1994} and a number of other observations. These issues are covered in detail in our previous work \citep[][]{ Kryshtal2019KPCB}.

Most solar flare models predict the appearance of DC electric fields that provide the simplest and most obvious mean to accelerate electrons. When a thermal plasma is under the influence of an electric field, a fraction of the electrons that have speeds greater than a critical speed, $v_c$, will be freely accelerated out of the thermal distribution \citep[]{Dreicer1959, Dreicer1960}. The critical speed is determined by the temperature of electrons, their number density, electric field strength and the resistivity of the plasma. \citet[]{Holman1985} estimated that the maximum number of electrons that can be accelerated is given by
\[
N\leq 4.3\times 10^{29}\left(\frac{B}{100 mT}\right)\frac{A}{5.3\times 10^{17} cm^2}\left(\frac{10^7 K}{T}\right)^{1/2},
\]
where $A$ is the area of the current sheet that is associated with the process of magnetic reconnection during a flare process.

To evidence the presence of large-scale quasi-static weak electric fields in the solar atmosphere \citet{Foukal1991} proposed using additional Stark broadening \citep[][]{Griem1974} of Balmer lines $H_{\beta}$ with $N\ge8$. Indeed, their proposition has now been confirmed, and several indirect evidence of electric current heating the plasma via Joule dissipation have been adduced. These evidences include the additional footpoint emission of flare loops observed during microwave spikes; extended thermal sources clearly seen in loops at coronal heights; a correlation between the shear of photospheric magnetic field with enhanced coronal heating and, many others \citep[see, e.g.][]{Benka1994}. Accordingly, a theoretical model has been developed by \citep[see, e.g.][]{Kryshtal2012, Kryshtal2014a} to show how small-scale plasma waves can be generated in pre-flare AR loops structures due to the presence of a weak background electric field. Such fields are known as sub-Dreicer fields if their amplitudes are much less than the amplitude of the local Dreicer field \citep{Miller1997}. The importance of such fields resides in the fact that the existence of sub-Dreicer fields in a loop can change the evolutionary process of a flare. Sub-Dreicer fields have successfully explained the hard X-ray spectra that is observed during flares, however they cannot explain the existence of very energetic electrons due to the finite length of small electric field. More importantly these weak electric fields are responsible for the production of runaway electrons. Direct sub-Deicer fields parallel to the magnetic field resulting from reconnecting field lines are possible candidates to explain the impulsively accelerated electrons \citep[see e.g.][]{Kuijpers1981, Holman1985, Moghaddam1990,Miller1997,Xu2007,Guo2013,Tsap2017,Zaitsev2018}

According to \citet{Heyvaerts1977}, the flaring process in an AR (loops, in particular) starts during a preheating phase, which precedes an impulsive phase and the plasma becomes visible in soft X-rays and Extreme Ultraviolet (EUV) wavelengths. The preheating phase begins as a result of the Buneman instability \citep[see, e.g.][]{Alexandrov1984, Chen1984}, that appears when the electron drift velocity exceeds the electron thermal velocity. Therefore, sub-Dreicer electric fields play an important role in the excitation of various types of instabilities and hence in the generation of small-scale plasma turbulence well before the preheating phase. The amplitudes of generated plasma waves during the linear stage of the initial perturbation do not exceed the level of thermal noise, but undamped waves can result in appearance of different types of three-wave interactions \citep[see, e.g.][]{Kryshtal2004, Kryshtal2012}. This aspect might influence the process of three-wave interaction and can be used for the short-term flare prediction \citep[][]{Aurass1994}.
Among the great number of waves that can appear in a magneto-active plasma \citep[][]{Alexandrov1984}, two in particular are of great interest, i.e. the kinetic Alfv\'{e}n wave (KAW) and the kinetic ion-acoustic wave (KIAW). Both these types of waves have unique properties \citep[see, e.g.][]{Hasegawa1976a} and can effectively accelerate particles, \citep{Miller1997}, take part in processes of three-wave interaction \citep[see e.g.][]{Hasegawa1976b, Brodin2006}, support the development of turbulence (which is a necessary element in the process of formation of pre-flare current sheets) and can cause abnormal resistivity \citep[][]{Somov1994}. In solar plasma studies all these processes and effects have been investigated in the post-flare stage of the plasma, without the presence of background electric field, i.e. when beams of energetic particles are supposed to be the main driver of KAW's instability \citep[see, e.g.][]{Zharkova2011b, Kumar2016, Druett2017}. On the other hand a large number of studies have been devoted to the study of KAW and plasma wave generation in solar atmosphere and solar flares, e.g. \citet[]{Voitenko2002, Wu2007, Tsiklauri2011, Artemyev2016}, etc.

The use of semi-empirical models of the flare chromosphere for AR forecasting may significantly increase its accuracy. However, this is possible only by taking into account a number of conditions, since the majority of these models describe the post-flare atmosphere. Therefore, only model which represent the quiet Sun atmosphere, e.g VAL (F) type \citep[see e.g][]{Vernazza1981} can be used to describe instabilities and corresponding wave generation in a pre-flare plasma. The use of MAVN (F1) \citep[][]{Machado1980} and FAL (PM) \citep[][]{Fontenla1993} models, previously considered by \citet{Kryshtal2019KPCB} for short-term flare prediction are applicable only under rare conditions, for example, by taking into account the special role of energetic particle beams \citep[see, e.g.][]{Harra2001}. The generation of instabilities may be also suppressed by secondary beams and, therefore, the beam models may play only additional role.

The present paper is devoted to the study the pre-flare stage of the plasma by using VAL (F) model, when the sub-Dreicer electric field acts as the main driver of the plasma instability that will drive KAWs and KIAWs without the need of energetic particle beams near loops' footpoints. Our approach follows closely the methodology developed recently by \citet[][]{Kryshtal2019KPCB}. Our study can also be applied to the situation when flares occur in succession in the same AR, i.e. when the percentage of energetic particles after the first flare has become too small to influence the onset of the succeeding flare \citep[see, e.g.][]{Harra2001}.
 
\section{Pre-flare plasma model and dispersion relation} \label{sec:ModelOfpre-flare}

It is well known that in a purely hydrogen plasma that is under the influence of an electric field, $E_0$, (with field strength less than the Dreicer field, $E_D$) parallel to the ambient magnetic field, a fraction of electrons with velocities greater than the critical value $(E_c/E_0)^{1/2}v_{Te}$ can overcome the drag force due to Coulomb collisions and become runaway electrons. Here $v_{Te}=\sqrt{k_B T_e/m_e}$ is the electron thermal speed, with $k_B$ the Boltzmann constant, $T_e$ the electron temperature, $m_e$ is the electron mass and the critical electric field is defined as 
\[
E_c\approx 2E_D=ek_D^2\Lambda_c=2\times 10^{-13}\frac{N(cm^{-3})\ln \Lambda_c}{T_e(eV)}.
\]
In the above expression $e$ is the electron charge, $k_D=\omega_{pe}/v_{Te}$ is the Debye wavelength, $\omega_{pe}$ is the electron plasma frequency, $N$ is the electron density, $T_e(eV)$ is the electron temperature measured in eV, and $\ln \Lambda_c$ is the Coulomb logarithm estimated at the critical velocity. As the runaway electrons are further accelerated by the electric field, more electrons are introduced to the runaway region
by collisions. As a result, the distribution function of electrons becomes distorted (with a runaway tail). We assume  
\begin{equation}
\left(\varepsilon_\mathrm{R}\right)_{min}\ll\varepsilon_\mathrm{R}\equiv\frac{E_{0}}{E_\mathrm{D}}\ll\left(\varepsilon_\mathrm{R}\right)_{max}\ll 1. \nonumber
\end{equation}
By taking into account the model assumptions presented by \citet[][]{Kryshtal2019KPCB} we consider the case when $T_e/T_i>1$, $\nu_{ei}\ne 0$ and $\varepsilon_{R}\ne0$. Here, $T_{i}$ is the ion temperature, $\nu_{ei}$ is the frequency of ion-electron collisions given by \citet{Alexandrov1984, Chen1984} and $\varepsilon_\mathrm{R}$ is the reduced amplitude of external weak large scale electric field in the local Deicer field units. Assuming that all perturbations oscillate with the same frequency, $\omega$, the linearised governing equations can be combined into the modified dispersion relation (MDR) of the form 
\begin{equation}
	\sum_{i=0}^{4}P_i\Omega^i=0, \label{Eq.31}
\end{equation}
where the coefficients $P_i$ are given by \citet[][]{Kryshtal2019KPCB}, $\Omega\equiv\omega/k_z V_A$ is the dimensionless frequency, $V_{A}$ is the Alfv\'{e}n speed, and $k_z$ is the longitudinal wave vector component (along the fields $\mathbf{E}_0\parallel\mathbf{B}_0$). The MDR takes into account spatial inhomogeneity of plasma density and temperature, paired Coulomb collisions and the existence of sub-Dreicer electric field. 

We are going to disregard all aperiodic roots of Eq. (\ref{Eq.31}). Finding physically acceptable solutions requires a very strict set of constraints on possible values of the plasma parameters. We are interested in weak amplification of waves, therefore the roots to the polynomial equation (\ref{Eq.31}) will have to satisfy the condition \citep[see][for details]{Kryshtal2019KPCB}
\begin{equation}
\Gamma_k\equiv\left.\frac{\gamma}{\omega}\right|_{\omega=\omega_k}\ll1, \, \left(k=1,....,4\right), \label{Eq.40}
\end{equation}
where the index $k$ labels the possible roots of the equation (\ref{Eq.31}), $\omega$ and $\gamma$ are the real and imaginary parts of the roots, respectively. This equation provides a measure of correctness for the use of linear approximation for the determination of growth rate of instability. Previous studies by \citet[][]{Kryshtal2004, Kryshtal2005b} determined and discussed the growth rate corresponding to the cases  $\mathbf{\bigtriangledown}n\ne0$; $\mathbf{\bigtriangledown}T=0$  and  $\mathbf{\bigtriangledown}n=0$; $\mathbf{\bigtriangledown}T\ne0$. Clearly, the case we are studying here ($\mathbf{\bigtriangledown}n\ne0$; $\mathbf{\bigtriangledown}T\ne0$) is the most general case, and it should not be considered as a linear combination of the above two simpler limits. 

The growth rate for the model can be written as \citep[see, e.g.][where the expressions of the functions $F_1$ and $F_2$ are also provided]{Kryshtal2019KPCB}
\begin{equation}
\Gamma_k\equiv\left.\frac{\gamma}{\omega}\right|_{\omega=\omega_k}=\sqrt{\frac{\pi}{2}}\frac{\left(\Omega-\beta_A \varepsilon_{R}\right)}{\beta \Omega^2}\left.\frac{F_1}{F_2}\right|_{\omega=\omega_k}\ll1, \, \left(k=1,....,4\right), \label{Eq.41}
\end{equation}
where
\begin{equation}
\beta_A\equiv\frac{V_{Te}}{V_A}. \nonumber
\end{equation}
and $\beta$ is the is plasma-beta parameter.

When determining the realistic solutions of our dispersion relation we took into account a few aspects that are driven by the physical background of the problem. First, in order to determine where the growth rate is positive (i.e. solutions are unstable), we excluded the solutions of the MDR that did not satisfy the criterion given by Eq. (\ref{Eq.40}). This restriction is equivalent with not considering fast aperiodic processes of amplification or damping of the waves. Next, from all possible solutions we considered only those which exhibit "sign-changed" growth rates. We considered the existence of the dividing curve $\Gamma_k=0$ on the surface $\Gamma_k=\Gamma\left(z_i, k_{*}\right), \, \left(k=1,....,4\right)$ as possible evidence for the appearance of undamped waves with small amplitudes in pre-flare plasma. Importantly, from an observational point of view, these amplitude values can exceed the level of thermal noise. Such appearance may be important for the existence of different three-wave interactions in pre-flare plasma and thus forecasting flares in AR.
	
\section{Investigation of kinetic and EM wave generation and stability} \label{sec:InvestigationOfStability}

In an earlier paper \citet[][]{Kryshtal2017} studied the modified dispersion relation (MDR) for low-frequency kinetic waves and the wave solutions were identified and classified for magnetic fields in the range $\mathbf{B}_0\in\left(\sqrt{1000}, 1000\right)$G, taking into account the influence of the sub-Dreicer electric field, drift plasma motions, paired Coulomb collisions and small-scale Bernstein turbulence when no complete inhomogeneity (density and temperature) was considered. In the same study we identified the combination of plasma parameters that correspond to the loop's footpoint in the chomosphere using the VAL (F) model \citep[]{Vernazza1981} at the height of $h=2.429$ Mm above the solar surface. These plasma parameters are shown in Table \ref{Tb.1}, and they will be used in our calculations.

\begin{table}
\caption{Main physical characteristics of a pre-flare plasma at the chromospheric part of a loop in a solar AR. Here $\omega_{*}$ is the ratio of plasma frequency, $\omega_{p}$, to the ion-cyclotron frequency, $\Omega_i$; $\nu_c\equiv\sigma\nu_{ei}/\Omega_i$ is the Coulomb collision frequency; $r_i$ is the ion-cyclotron gyroradius, $L$ is the density characteristic scale of plasma spatial inhomogeneity. All these values were extracted from the VAL (F) solar atmospheric model. Here we assume that the temperature ($T$) and number density ($n$) of ions and electrons are equal. The quantity $\sigma$ in the expression for $\nu_c$ plays the role of the free parameter and physically bounded as 
$ 1\le\sigma\le\sigma_{max}$. $\sigma=1$ corresponds to hypothetical situation when the condition for instability development is maximal, i.e. when ion-electron collisions completely dominate. The value $\sigma=\sigma_{max}$ corresponds to the case when the instability is completely suppressed by the collisions. It has been estimated by, e.g. \citet{Kryshtal2002a} and \cite{Kryshtal2012}, that in this limiting case $\sigma_{max}<10$.}
\label{Tb.1}
\begin{tabular}{ccccccccc}     
  \hline                   
 $n$, & $T$, & h & $|\mathbf{B}_0|$, & $\omega_{*}$ & $\nu_c$ & $\beta$ & $r_i$, & $L$, \\
 $10^9cm^{-3}$ & $10^5 $$K$ & km & mT &  & $10^{-4}$ & $10^{-4}$ & cm & $10^5$ cm \\
  \hline
 2.57 & 4.47 & 2429 & 3.162 & 5.15 & 4.61 & 3.64 & 4.66 & 1\\
  \hline
\end{tabular}
\end{table}

\begin{table}
\caption{The calculated values of the main characteristics of instability of the modified KIAW (the $\Omega_3$ and $\Omega_4$ roots of the modified dispersion equation) and KAW (the $\Omega_2$ root) for the plasma parameters given in Table \ref{Tb.1} for the VAL (F) solar atmospheric model. Here $\nu_{turb}$ describes the saturated Bernstein turbulence.}
\label{Tb.2}
\begin{tabular}{cccc}     
\hline                   
Wave type & $\left(k\right)_{bound}$ & $\left(z\right)_{bound}$ & $R_{0\rho}$ \\
  \hline
direct KIAW $\left(\nu=\nu_{turb}\right)$ & 0.029 & 0.011 & -1.57\\
inverse KIAW $\left(\nu=\nu_{turb}\right)$ & -0.0172 & 0.0604 & -1.83\\
inverse KAW $\left(\nu=\nu_{turb}\right)$ & -0.0132 & 0.011 & -1.48\\
inverse KIAW $\left(\nu=\nu_{c}\right)$ & -0.0035 & 0.125 & -2.55\\
  \hline
\end{tabular}
\end{table}

In this paper we use only one value of magnetic field strength $|\mathbf{B}_0|$ for each model of solar atmosphere. Throughout our calculations we are going use the value of $\sigma=5$ for the free parameter of the model \citep[see e.g.][]{Kryshtal2012}, which allows us to take into account, on the phenomenological level, the contributions of all kinds of mutual collisions of charged particles within the BGK integral. Furthermore, we assume that paired Coulomb collisions are the dominant effect in the pre-flare plasma prior to the turbulence appearance. When Bernstein turbulence begins to dominate (the turbulent phase), the collision rate is described by the effective collision frequency, $\nu_{eff}$, \citep[][]{Galeev1972}, valid if $\nu_{eff}\gg \mu\Omega_i$ (here, $\mu\equiv\sqrt{m_e/m_i}$ is electron-to-ion mass ratio) and its value is given by  
\begin{equation}
\nu_{eff}=\frac{\Gamma}{z_e}\Omega_e, \label{Eq.45}
\end{equation}
where $\Gamma$ is the linear reduced growth rate of second quasi-Bernstein harmonics \citep[see, e.g.][]{Kryshtal2012}, $\Omega_e$ is the electron cyclotron frequency and $z_e$ is the electron kinetic parameter
\begin{equation}
z_e\equiv \frac{k_{\perp}^2V_{Te}^2}{\Omega_e^2}. \label{Eq.46}
\end{equation}

The problem of the influence of small-scale Bernstein turbulence on low-frequency waves is a very important aspect of our problem and it has been previously investigated by \citet[]{Kryshtal2017}. Their analysis showed that second harmonics of oblique quasi-Bernstein modes are possible in the range of magnetic field that is characteristics of pre-flare AR chromospheres (from few units of mT up to few tens of mT). The dispersion relation they derived admitted analytical solutions that describe the KAW and KIAW 
\begin{equation}
\Omega_{KAW}=\pm\sqrt{1+z_i\left(t_{*}+\frac{3}{4}\right)}, \label{Eq.49}
\end{equation}
\begin{equation}
\Omega_{KIAW}=\pm\mu\beta_A\sqrt{\frac{1-z_i}{1+z_i t_{*}}}. \label{Eq.50}
\end{equation}
where the $\pm$ sign stands for forward and backward propagating waves, $z_i$ is the ion kinetic parameter defined by
\begin{equation}
z_i\equiv \frac{k_{\perp}^2V_{Ti}^2}{\Omega_i^2}, \nonumber
\end{equation}
with $V_{Ti}$ being the ion thermal velocity, $k_{\perp}^2\equiv k^2-k_z^2$ the perpendicular component of wave vector, $k_z$ the longitudinal component (along the fields $\mathbf{E}_0\parallel\mathbf{B}_0$), and $t_{*}=T_e/T_i$ is the ratio of electron and ion temperatures.

We should mention that the nature of the solutions should not be different from the case studied by \citet[]{Kryshtal2017}, however, due to the fact that the collisions are described by $\nu_{eff}\ne0$, we expect that the solutions of the MDR (i.e. the roots of the fourth order polynomial given by Eq. \ref{Eq.31}) will be modified in magnitude due to the presence of extra physical effects. That is why, the solutions studied here will be modifications of results obtained earlier by \citet[]{Kryshtal2017}.

At the chromospheric height shown in Table~\ref{Tb.1}, using the semi-empirical solar atmospheric VAL (F) model \citep[][]{Vernazza1981}, we have investigated the stability of the four roots of the MDR (Eq.\ref{Eq.31}). Our analysis show that the four roots have different behaviour and they represent the forward and backward propagating modified KAW, and modified KIAW, respectively. Our analysis has been repeated for six different values of magnetic field strength, i.e. $|\mathbf{B}_0|=\sqrt{1000}\approx 31.62$mT; 100mT; 300mT; 500mT; 700mT and 1000mT. Although not all values are strictly adequate for chromospheric conditions, we used this extended magnetic field strength interval to ensure robustness of our results. In addition, solutions to the dispersion relation have been sought so that the parameters of the problem were varied in the intervals
\begin{eqnarray}
&& 5\cdot10^{-7}\le\varepsilon_{R}\le10^{-2};  \nonumber \\
&& -5\le\delta\le5;\,\,\left(\delta=\pm\frac{L}{L_T}\right);  \nonumber \\
&& 1\le\sigma\le 6;  \label{Eq.51} \\
&& 0.001\le k_{*}\le 0.25;  \nonumber \\
&& 0.001\le z_i \le 0.25.  \nonumber 
\end{eqnarray}
Here, $L_T$ is the characteristic scale of inhomogeneity of plasma temperature, $k_{*}=k_{z}/k$. The boundary values of the above parameters are the values at which an instability appears and begins to develop, i.e. when the imaginary part of the frequency changes sign from negative to positive. 

Numerical analysis reveals that for the model we are using, the boundary values that satisfy the imposed physical restrictions occur when $\left(t_{*}\right)_{bound}=4$ and $\sigma_{max}=6$, i.e. instabilities with a growth rate that is changing sign satisfying the condition (\ref{Eq.40}), appear when $t_{*}=4$ and disappear when $\sigma=6$. In our calculations the value of $\delta$ was chosen such that it leads to the most favourable instability development and closest (in absolute value) to the classical value of $\delta=2$ corresponding to the drift-resistive instability \citep[see, e.g.][]{Somov1994}. 

The boundary values for $\varepsilon_{R}$, $k_{*}$ and $z_i$ correspond to the value $\Gamma_{1+}$ - the first positive value of the reduced growth rate in the half-space $\Gamma>0$ after the growth rate has changed its sign from negative to positive. It is easy to show that all terms in the expression of the reduced growth rate given by Eqs. (\ref{Eq.41}), describing the effect of the drift motions contain the factor
\begin{equation}
\rho\equiv\frac{\rho_i}{L}. \label{Eq.54}
\end{equation}
In order to estimate the influence of the spatial inhomogeneity in plasma temperature and density on the reduced growth rate, we define the parameter
\begin{equation}
R_{0\rho}\equiv \frac{\Gamma_0-\Gamma_{\rho}}{\Gamma_0}, \label{Eq.55}
\end{equation}
where $\Gamma_0\equiv\Gamma_{1+}$ when $\rho=0$ and $\Gamma_{\rho}\equiv\Gamma_{1+}$ when $\rho\ne 0$. 

All the most significant boundary values for the parameters of low-frequency instabilities, which we have investigated for the VAL (F) atmospheric model are presented in Table~\ref{Tb.2}. For the particular case we are investigating we assumed that $\left(t_{*}\right)_{bound}=4$, $\sigma=5$, $\delta=-2$ and $10^{-4}\le\left(\varepsilon_{R}\right)_{bound}\le 1.2\times 10^{-4}$. In this case, three out of four possible wave types can be generated \citep[][]{Kryshtal2017KPCB}. It should be noted that Table~\ref{Tb.2} contains data for the instabilities which show a sign-change, i.e. when the reduced growth rates given by Eq. (\ref{Eq.41}) pass through their zero. 

In ARs, at the foot-points of magnetic loops  (at heights which correspond to the low-middle chromosphere), the generation of second harmonics of Bernstein modes are possible as a result of corresponding instability with has anomalously low threshold of excitation and, therefore, can be transformed into the state of saturated turbulence \citep[see e.g.][]{Kryshtal2019KPCB}. Hence, the appearance and development of kinetic waves, which have much higher threshold of excitation, may be possible on the background of small-scale saturated Bernstein turbulence.
 At this stage, Bernstein modes can generate undamped waves (this has been proved in previous studies by \citet[][]{Kryshtal2012, Kryshtal2015}. In such a situation the appearance of an EM wave, - a spike of microwave emission, - the forerunner of a flare, is possible through a nonlinear three-wave interaction \citep[][]{Willes1996}:
\begin{equation}
B_1+B_2\rightarrow EMW, \label{Eq.56}
\end{equation}
where $B_1$ and $B_2$ represent the Bernstein modes and EMW stands for the transverse electromagnetic wave. In their study \citet[][]{Willes1996}  have shown that in the case of the above wave interaction the probability of the process is maximal if the $B_1$ and $B_2$ modes belong to the same morphological class. However, the process of coalescence has an extremely small probability. Therefore, here we investigate the more complicated, yet more realistic outcome for the Bernsteinian phase of the instability, when it transforms into a small-scale turbulence, then reaches a state of saturation, in which kinetic waves will be generated. These kinetic waves (KAW and KIAW) are the necessary participants in a number of three-wave interactions \citep[see, e.g.][]{Yukhimuk1998, Yukhimuk2000} that, in principle, can produce the EM forerunner of a flare or can results in the depolarisation of solar radio emission \citep[see e.g.][]{Sirenko2000}. Usually  the probability of decay in three-wave interaction is supposed to be much more probable than coalescence. Obviously, in the case when one of the waves participating in this process (or both) is propagating with no damping, this probability can only increase.

In a previous study \citet[][]{Kryshtal2015} investigated the generation and development of the instability of the first harmonic of quasi-Bernstein modes, which actually coincides with upper-hybrid wave (UHW) in the context of our current model. \citet[][]{Yukhimuk1998} have proved that electromagnetic emission in the microwave range can be produced through the process of three-wave interaction of the type
\begin{equation}
UHW\rightarrow KAW+EMW. \label{Eq.57}
\end{equation}
\citet[][]{Kryshtal2015} have shown that first harmonic of quasi-Bernstein modes can produce small amplitude undamped waves. It is clear, that in the case when the same waves are generated through the process of KAW's instability development, three-wave processes governed by Eq. (\ref{Eq.57}) become more probable EM waves. The same authors estimated the frequency of the EM emission and have compared it with emission from observations. Other comparisons with observations by, e.g. \citet{Bogod2009}, \citet{Zharkova2011b} and \citet{Kashapova2012} have shown good, if not exact agreement with theory. We have to emphasise that kinetic waves are necessary participants in all known plasma processes of nonlinear three-wave interactions, not only because they are possibly responsible for generating the spikes of microwave emission, but also in the mutual transformation of one wave to another \citep[][]{Hasegawa1976a, Hasegawa1976b, Kryshtal2014c, Lyubchyk2014}, e.g.
\begin{equation}
KAW\rightarrow KAW^{\prime}+KIAW. \label{Eq.58}
\end{equation}
The above three-wave interaction scheme means that generation of one kinetic Alfv\'{e}n wave may lead to the generation of the second one as a result of the cascade process.
\begin{eqnarray}
KIAW\rightarrow KIAW^{\prime}+KAW, \label{Eq.58a} \\
KAW\rightarrow KAW^{\prime}+KAW^{\prime \prime}. \label{Eq.58b}
\end{eqnarray}
Given their properties, kinetic waves can be an excellent candidate for plasma heating, they can depolarize solar EM emission \citep[]{Sirenko2000}, and can accelerate particles in pre-flare ARs \citep[]{Aschwanden2005}. Thus, the possibility of generation of kinetic waves (KAW and KIAW) on the background of small-scale Bernstein turbulence, which we have shown, seems to be very important, because it gives evidence that small-scale turbulence cannot destroy the process of instability development and the time of generation of kinetic waves may be long enough for a number of three-wave processes to be realised.

\begin{figure}
\centering
\mbox{\includegraphics[scale=0.07, angle=0]{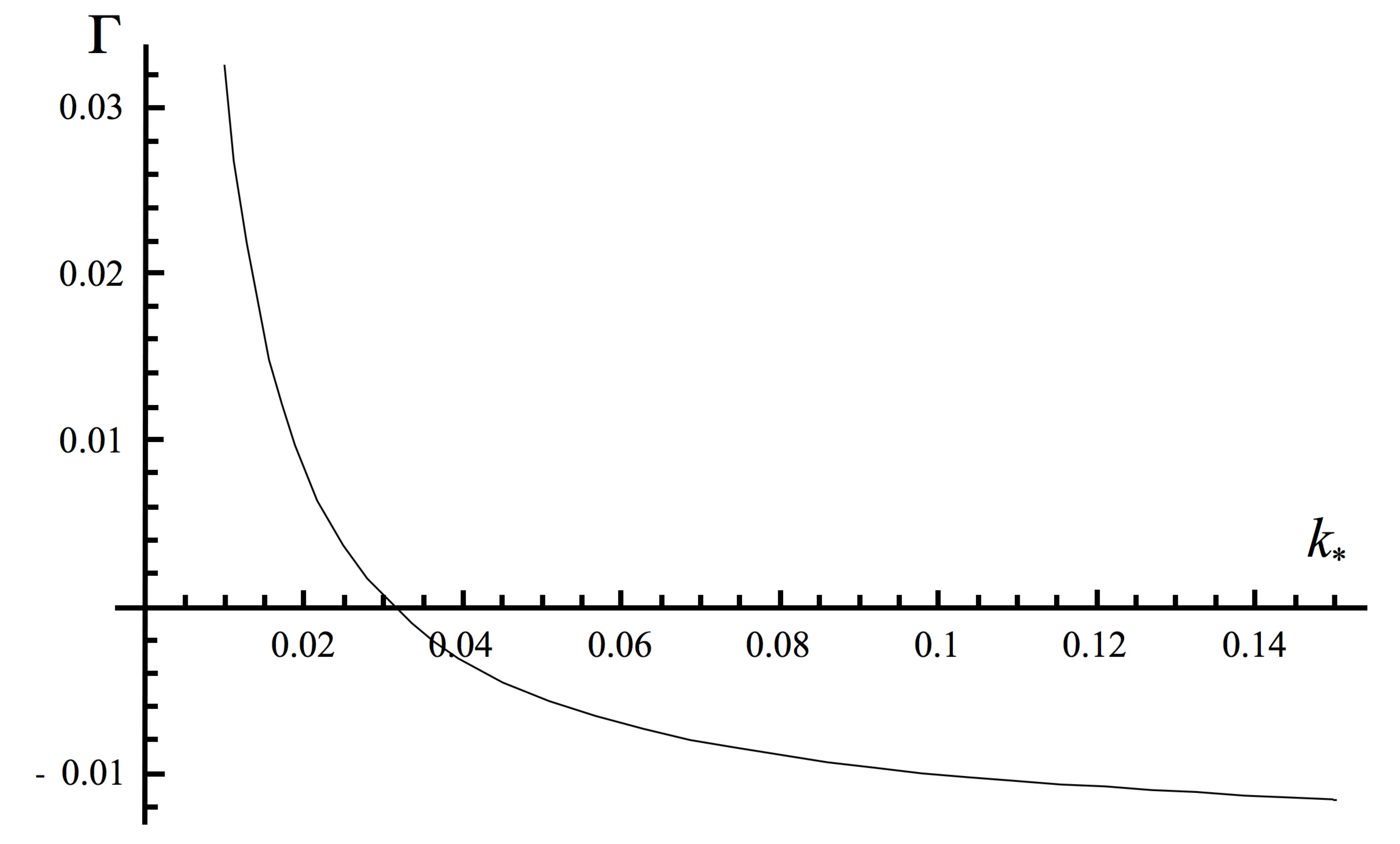}} \mbox{\includegraphics[scale=0.07, angle=0]{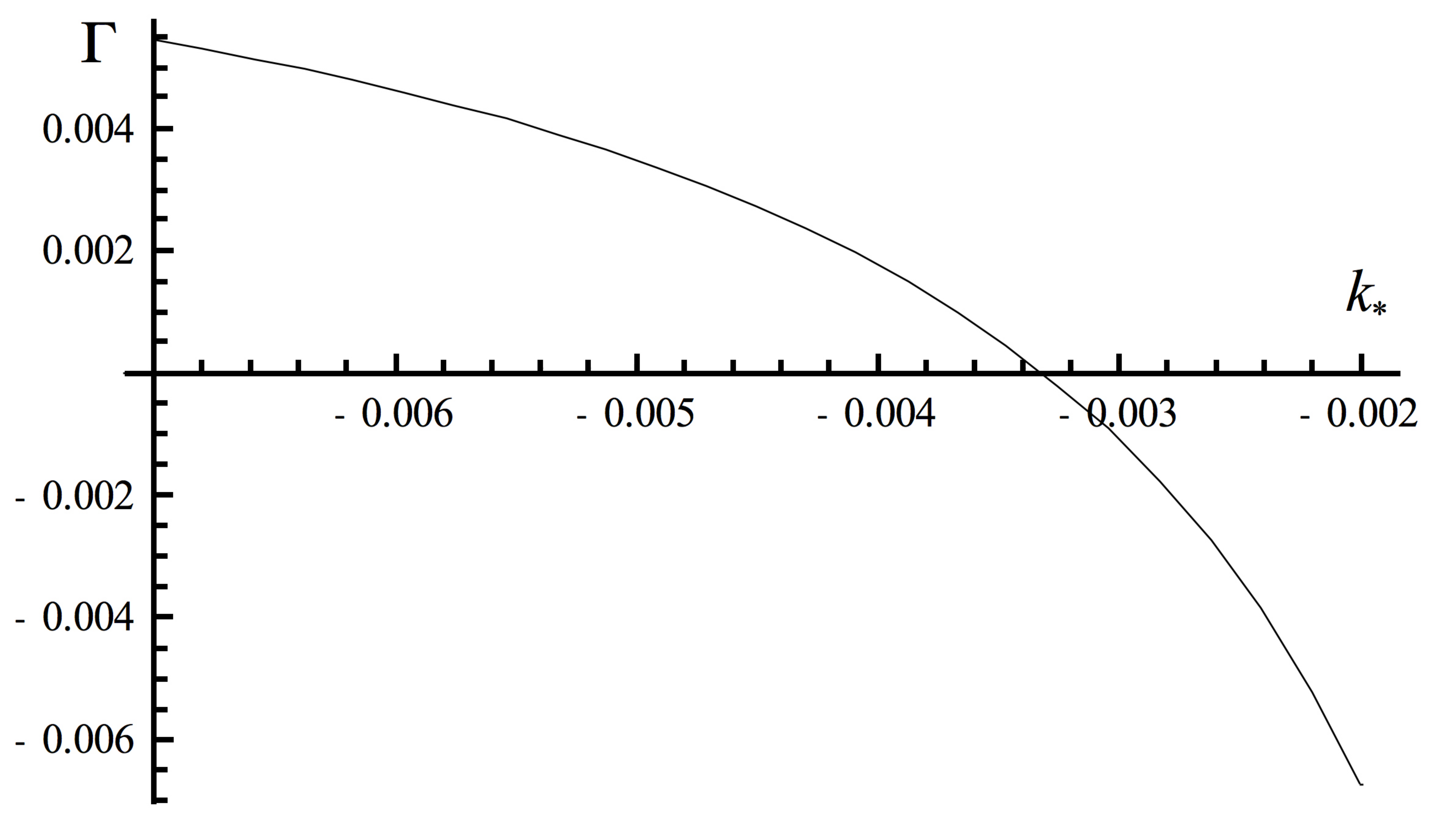}}
\mbox{\includegraphics[scale=0.07, angle=0]{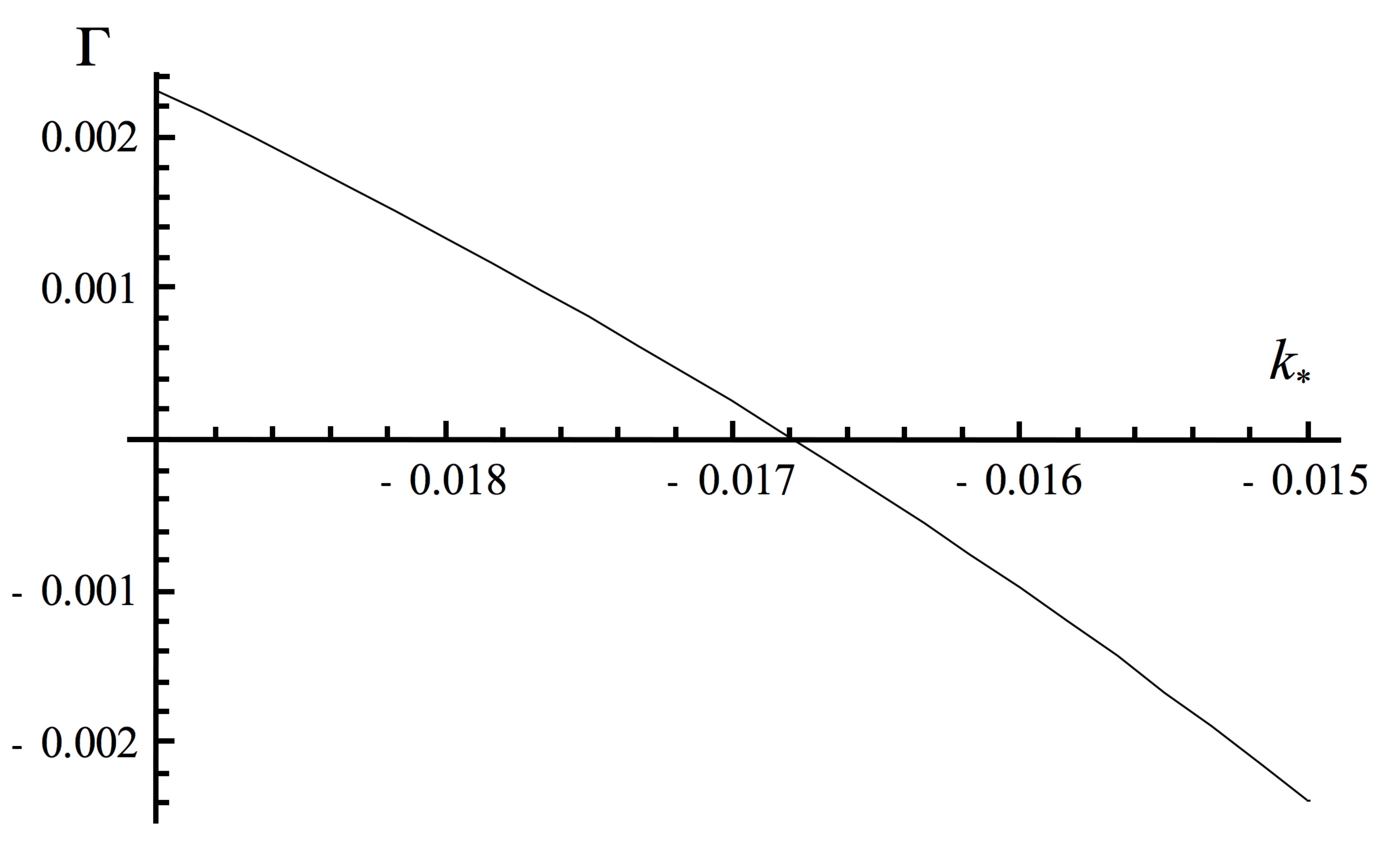}} \mbox{\includegraphics[scale=0.07, angle=0]{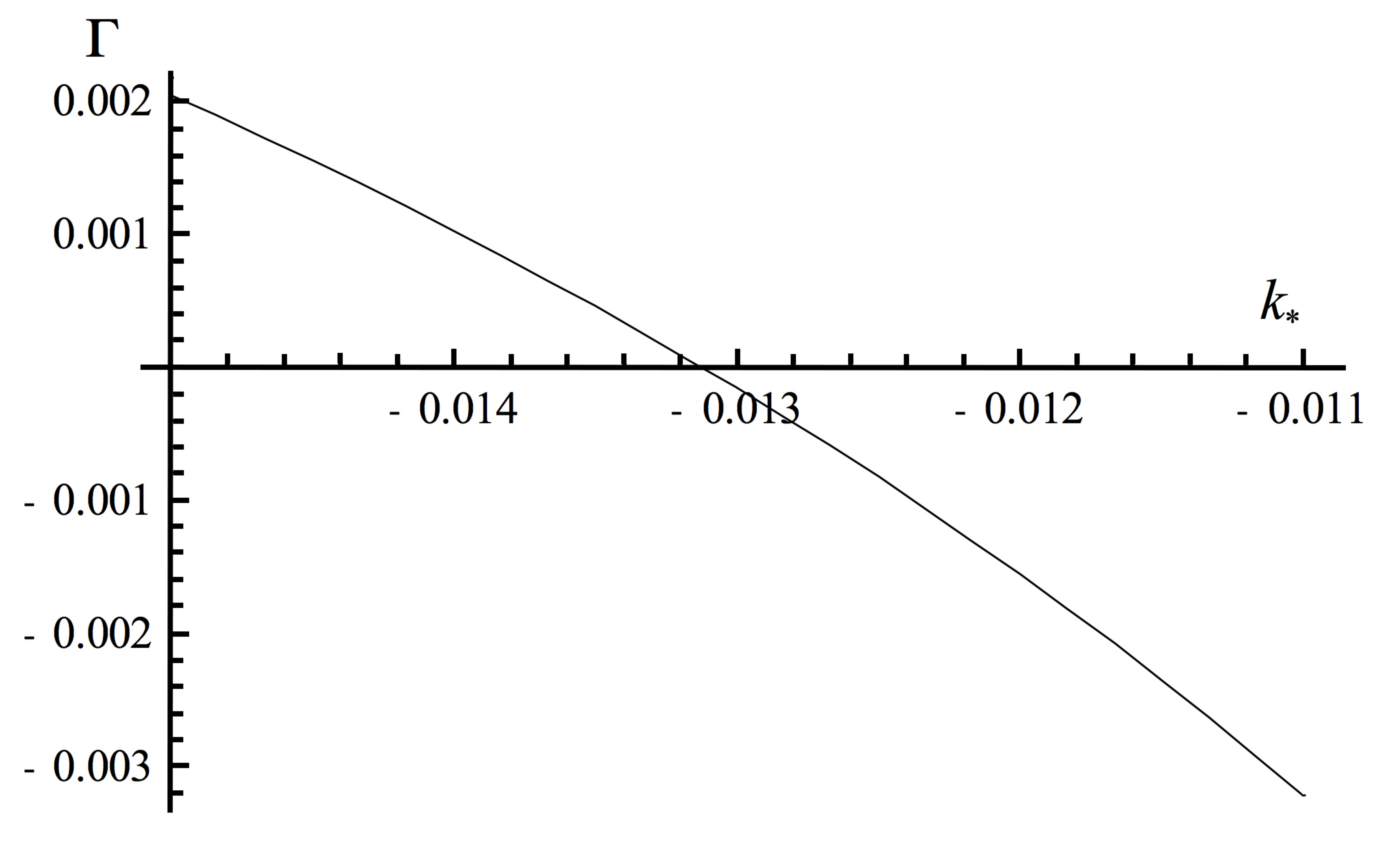}}
\caption{The reduced growth rates of instability, $\Gamma$, for a VAL (F) solar atmospheric model as a function of $k_{*}=k_{z}/k$. The four different cases shown here are: direct KIAW (top left panel); inverse KIAW (top right) for the plasma with Coulomb conductivity; inverse KIAW (bottom left) and inverse KAW (bottom right), both for the plasma with small-scale Bernstein turbulence.}
\label{fig:12}
\end{figure}

\section{Results and discussion} \label{sec:ResultsAndDiscussion}

Our study was devoted to the analysis of the excitation and development of low-frequency instabilities in fully ionised pre-flaring chromospheric plasma. The plasma parameters (density and temperature) were given by the  VAL (F) solar atmospheric model. 

In Figure 1 (top left panel) we plot the reduced growth rate of instability, $\Gamma$, as a function of $k_{*}$ for the considered VAL (F) solar atmospheric model. In this case the reduced values of effective turbulence and Coulomb collision frequencies satisfy the inequality
\begin{equation}
\nu_{turb}\gg\nu_{c}, \label{Eq.59}
\end{equation}
The value of $R_{0\rho}$ demonstrates that corrections to the induced growth rate $\Gamma$ that take into account the spatial inhomogeneities of plasma temperature and density may be negligibly small in absolute values, but the relative change of $\Gamma$ may be considerable enough.

As it can be seen from the Figure 1, among the four roots of the modified dispersion relation, only three (direct/inverse KIAW and inverse KAW for the plasma with Coulomb conductivity and small-scale Bernstein turbulence) are unstable and have alternating signs. The root which corresponds to the direct KAW is stable. Therefore, there is a fundamental difference between the present study and \citet[][]{Kryshtal2019KPCB}, where the VAL (F) model does not include the unstable roots of the modified dispersion relation. This result was obtained with complete coincidence of the entire “calculated part” (i.e, by taking into account similar approximations and equations presented in \citet[][]{Kryshtal2019KPCB}, but by applying different values of the parameters of the wave disturbance. The physical parameters we worked with can be considered as typical for the  pre-flare stage and instabilities which may appear in plasma during the pre-flare phase are non-beam instabilities. 

Our results show that in the current circuit of the loop near its foot-point in the active region, the generation of kinetic waves is possible as a result of the development of an instability during flare process, when electron thermal velocity is greater then current velocity, that is before appearance of Buneman instability and beginning of the preliminary heating phase. For semi-empirical solar atmosphere models (including the VAL (F) model) there is an area at the magnetic loop foot-point where instability threshold for the second quasi-Bernstein harmonic is very low \citep{Kryshtal2012, Kryshtal2017}. 

We have shown that the development of instability of kinetic wave is possible not only in a plasma with Coulomb conductivity, but also in the presence of Bernstein turbulence, i.e. this turbulence will not destroy the generation of kinetic waves. Therefore, waves may have enough time for participation in the three-wave interactions, which may generate electromagnetic wave \citep{Yukhimuk1998}. Observation of these waves can be considered as necessary condition for short term flare forecast in AR.

Using standard methods outlined in the paper we derived a dispersion relation in the form of a quartic polynomial whose roots represent two KAWs and two KIAWs propagating in opposite direction. Low-frequency instability can spontaneously develop when paired Coulomb collisions dominate but also when the small-scale Bernstein turbulence is dominant. Among these roots (for the chosen plasma paramaters) only the direct KAW shows stable behaviour as can be seen from the Table \ref{Tb.2}. For the considered case we have found a slow instability with sufficient low threshold (but much higher than the threshold of excitation of the second quasi-Bernstein harmonic). The instability recovered here is important as it appears in the loop footpoint when traditional drivers of instabilities, i.e. beams of energetic particles are not present \citep[see, e.g.][]{Battaglia2014}.

Our calculations have shown that the spatial inhomogeneity of plasma temperature and density can influence the growth rate of the instability through the values of the drift frequency. Although the absolute value of changes in the reduced growth rate remain very small, the relative changes of the growth rate may be considerable enough to notably influence the solutions of the modified dispersion relation. 

The kinetic wave generated due to the instability was suitable enough to determine the influence of the drift plasma motion on the value of the growth rate, because the root corresponding to this wave has a growth rate which changes sign, and its first positive value in the half-space $\Gamma\left(z_i, k_{*}\right)>0$ is the best candidate for comparing the cases that correspond to the growth rate.

The problem of generation of the kinetic waves is important in studying the physical processes in a pre-flare plasma because these waves have their own electric field, which allows them, in principle, to play an important role in the pre-flare acceleration of the particles and plasma heating through the generation of turbulence and abnormal resistance. In contrast to MHD instabilities, which have been frequently observed in AR loops, kinetic instabilities and kinetic plasma waves involve spatial and temporal scales \citep[see, e.g.][]{Hasegawa1976a, Galeev1983, Galeev1985} that are impossible to observe with the current instruments. However, it has been shown that the generation of kinetic waves increases the probability of three-wave processes which will, in turn, produce EM waves which can be detected by observers and thus may become an early warning signal of high energy processes in a pre-flare AR \citep[][]{Farnik1998, Schmahl1986, Willes1996}. 

\citet[][]{Nunez2005} pointed out that the flare forecasting precision can be increased by combining synoptical and causal forecast, where the synoptical forecast is based on the morphology of ARs before a flare. The causal forecast is based on knowledge of the physical mechanisms of the flaring process. The results presented in this paper may be considered as only necessary part of the causal forecast.

\section{Acknowledgements} \label{sec:Acknowledgements} OC would like to thank the Ukrainian Scientific and Technical Center, PN 6060; Integrated Scientific Programme of the National
Academy of Science of Ukraine on Space Research. VF, GV and IB thank to The Royal Society, International Exchanges Scheme, collaboration with Chile (IE170301) and Brazil (IES$\backslash$R1$\backslash$191114). VF and GV are grateful to the Science and Technology Facilities Council (STFC) grant ST/M000826/1 for support provided. This research is also partially funded by the European Union’s Horizon 2020 research and innovation program under grant agreement No. 824135 (SOLARNET). This work also greatly benefited from the discussions at the ISSI workshop "Towards Dynamic Solar Atmospheric Magneto-Seismology with New Generation Instrumentation"

\bibliography{SOLA_Kryshtal_2020_R}{}

\begin{thebibliography}{}
\expandafter\ifx\csname natexlab\endcsname\relax\def\natexlab#1{#1}\fi

\bibitem[{Alexandrov(1984)}]{Alexandrov1984}
Alexandrov, A., B.~L. R.~A. 1984, Principles of Plasma Electrodynamics
  (Springer-Verlag)

\bibitem[{{Artemyev} {et~al.}(2016){Artemyev}, {Zimovets}, \&
  {Rankin}}]{Artemyev2016}
{Artemyev}, A.~V., {Zimovets}, I., \& {Rankin}, R. 2016, \aap, 589

\bibitem[{{Aschwanden}(2001)}]{Aschwanden2001}
{Aschwanden}, M.~J. 2001, \apj, 560, 1035

\bibitem[{{Aschwanden}(2005)}]{Aschwanden2005}
---. 2005, {Physics of the Solar Corona. An Introduction with Problems and
  Solutions (2nd edition)}

\bibitem[{{Aurass}(1994)}]{Aurass1994}
{Aurass}, H. 1994, in IAU Colloq. 144: Solar Coronal Structures, ed.
  V.~{Rusin}, P.~{Heinzel}, \& J.-C. {Vial}, 251--256

\bibitem[{{Battaglia} {et~al.}(2014){Battaglia}, {Fletcher}, \&
  {Sim{\~o}es}}]{Battaglia2014}
{Battaglia}, M., {Fletcher}, L., \& {Sim{\~o}es}, P. J.~A. 2014, \apj, 789, 47

\bibitem[{{Benka}(1994)}]{Benka1994}
{Benka}, S.~G. 1994, in Proceedings of Kofu Symposium, 225--229

\bibitem[{{Bogod} \& {Yasnov}(2009)}]{Bogod2009}
{Bogod}, V.~M., \& {Yasnov}, L.~V. 2009, \solphys, 255, 253

\bibitem[{{Brodin} {et~al.}(2006){Brodin}, {Stenflo}, \& {Shukla}}]{Brodin2006}
{Brodin}, G., {Stenflo}, L., \& {Shukla}, P.~K. 2006, \solphys, 236, 285

\bibitem[{{Charikov} \& {Shabalin}(2015)}]{Charikov2015}
{Charikov}, Y.~E., \& {Shabalin}, A.~N. 2015, Geomagnetism and Aeronomy, 55,
  1104

\bibitem[{{Chen}(1984)}]{Chen1984}
{Chen}, F.~F. 1984, {Introduction to Plasma Physics and Controlled Fusion,
  Volume 1: Plasma Physics} (Springer-Verlag)

\bibitem[{{Dreicer}(1959)}]{Dreicer1959}
{Dreicer}, H. 1959, Physical Review, 115, 238

\bibitem[{{Dreicer}(1960)}]{Dreicer1960}
---. 1960, Physical Review, 117, 329

\bibitem[{{Druett} {et~al.}(2017){Druett}, {Scullion}, {Zharkova}, {Matthews},
  {Zharkov}, \& {Rouppe van der Voort}}]{Druett2017}
{Druett}, M., {Scullion}, E., {Zharkova}, V., {et~al.} 2017, Nature
  Communications, 8, 15905

\bibitem[{{F{\'a}rn{\'\i}k} \& {Savy}(1998)}]{Farnik1998}
{F{\'a}rn{\'\i}k}, F., \& {Savy}, S.~K. 1998, \solphys, 183, 339

\bibitem[{{Fontenla} {et~al.}(1993){Fontenla}, {Avrett}, \&
  {Loeser}}]{Fontenla1993}
{Fontenla}, J.~M., {Avrett}, E.~H., \& {Loeser}, R. 1993, \apj, 406, 319

\bibitem[{{Foukal} \& {Hinata}(1991)}]{Foukal1991}
{Foukal}, P., \& {Hinata}, S. 1991, \solphys, 132, 307

\bibitem[{{Galeev} {et~al.}(1972){Galeev}, {Lominadze}, {Pataraya}, {Segdeev},
  \& {Stepanov}}]{Galeev1972}
{Galeev}, A.~A., {Lominadze}, D.~G., {Pataraya}, A.~D., {Segdeev}, R.~Z., \&
  {Stepanov}, K.~N. 1972, ZhETF Pisma Redaktsiiu, 15, 417

\bibitem[{{Galeev} \& {Sudan}(1983)}]{Galeev1983}
{Galeev}, A.~A., \& {Sudan}, R.~N. 1983, {Handbook of plasma physics. Vol. 1:
  Basic plasma physics I.}

\bibitem[{{Galeev} \& {Sudan}(1985)}]{Galeev1985}
---. 1985, {Handbook of plasma physics. Vol. 2: Basic plasma physics II.}

\bibitem[{{Griem}(1974)}]{Griem1974}
{Griem}, H.~R. 1974, {Spectral line broadening by plasmas}

\bibitem[{{Guo} {et~al.}(2013){Guo}, {Emslie}, \& {Piana}}]{Guo2013}
{Guo}, J., {Emslie}, A.~G., \& {Piana}, M. 2013, \apj, 766, 28

\bibitem[{{Harra} {et~al.}(2001){Harra}, {Matthews}, \& {Culhane}}]{Harra2001}
{Harra}, L.~K., {Matthews}, S.~A., \& {Culhane}, J.~L. 2001, \apjl, 549, L245

\bibitem[{{Hasegawa} \& {Chen}(1976{\natexlab{a}})}]{Hasegawa1976b}
{Hasegawa}, A., \& {Chen}, L. 1976{\natexlab{a}}, Physics of Fluids, 19, 1924

\bibitem[{{Hasegawa} \& {Chen}(1976{\natexlab{b}})}]{Hasegawa1976a}
---. 1976{\natexlab{b}}, \prl, 36, 1362

\bibitem[{{Heinzel} {et~al.}(2007){Heinzel}, {Dorotovi{\v{c}}}, \&
  {Rutten}}]{Heinzel2007}
{Heinzel}, P., {Dorotovi{\v{c}}}, I., \& {Rutten}, R.~J. 2007, {The Physics of
  Chromospheric Plasmas}, Vol. 368

\bibitem[{{Heyvaerts} {et~al.}(1977){Heyvaerts}, {Priest}, \&
  {Rust}}]{Heyvaerts1977}
{Heyvaerts}, J., {Priest}, E.~R., \& {Rust}, D.~M. 1977, \apj, 216, 123

\bibitem[{{Holman}(1985)}]{Holman1985}
{Holman}, G. 1985, \apj, 293, 584

\bibitem[{{Kashapova} {et~al.}(2012){Kashapova}, {Meshalkina}, \&
  {Kisil}}]{Kashapova2012}
{Kashapova}, L.~K., {Meshalkina}, N.~S., \& {Kisil}, M.~S. 2012, \solphys, 280,
  525

\bibitem[{{Kryshtal} {et~al.}(2015){Kryshtal}, {Fedun}, {Gerasimenko}, \&
  {Voitsekhovska}}]{Kryshtal2015}
{Kryshtal}, A., {Fedun}, V., {Gerasimenko}, S., \& {Voitsekhovska}, A. 2015,
  \solphys, 290, 3331

\bibitem[{{Kryshtal}(2002)}]{Kryshtal2002a}
{Kryshtal}, A.~N. 2002, Journal of Plasma Physics, 68, 137

\bibitem[{{Kryshtal}(2005)}]{Kryshtal2005b}
---. 2005, Journal of Plasma Physics, 71, 729

\bibitem[{{Kryshtal} \& {Gerasimenko}(2004)}]{Kryshtal2004}
{Kryshtal}, A.~N., \& {Gerasimenko}, S.~V. 2004, \aap, 420, 1107

\bibitem[{{Kryshtal} {et~al.}(2012){Kryshtal}, {Gerasimenko}, \&
  {Voitsekhovska}}]{Kryshtal2012}
{Kryshtal}, A.~N., {Gerasimenko}, S.~V., \& {Voitsekhovska}, A.~D. 2012,
  Advances in Space Research, 49, 791

\bibitem[{{Kryshtal} {et~al.}(2014{\natexlab{a}}){Kryshtal}, {Gerasimenko}, \&
  {Voitsekhovska}}]{Kryshtal2014a}
---. 2014{\natexlab{a}}, \apss, 349, 637

\bibitem[{{Kryshtal} {et~al.}(2014{\natexlab{b}}){Kryshtal}, {Gerasimenko},
  {Voitsekhovska}, \& {Cheremnykh}}]{Kryshtal2014c}
{Kryshtal}, A.~N., {Gerasimenko}, S.~V., {Voitsekhovska}, A.~D., \&
  {Cheremnykh}, O.~K. 2014{\natexlab{b}}, Kinematics and Physics of Celestial
  Bodies, 30, 147

\bibitem[{{Kryshtal} {et~al.}(2019){Kryshtal}, {Voitsekhovska}, \&
  {Gerasimenko}}]{Kryshtal2019KPCB}
{Kryshtal}, A.~N., {Voitsekhovska}, A.~D., \& {Gerasimenko}, S.~V. 2019,
  Kinematics and Physics of Celestial Bodies, 35, 105

\bibitem[{{Kryshtal} {et~al.}(2017{\natexlab{a}}){Kryshtal}, {Voitsekhovska},
  {Gerasimenko}, \& {Cheremnykh}}]{Kryshtal2017}
{Kryshtal}, A.~N., {Voitsekhovska}, A.~D., {Gerasimenko}, S.~V., \&
  {Cheremnykh}, O.~K. 2017{\natexlab{a}}, Kinematics and Physics of Celestial
  Bodies, 33, 149

\bibitem[{{Kryshtal} {et~al.}(2017{\natexlab{b}}){Kryshtal}, {Voitsekhovska},
  {Gerasimenko}, \& {Cheremnykh}}]{Kryshtal2017KPCB}
---. 2017{\natexlab{b}}, Kinematics and Physics of Celestial Bodies, 33, 149

\bibitem[{{Kuijpers} {et~al.}(1981){Kuijpers}, {van der Post}, \&
  {Slottje}}]{Kuijpers1981}
{Kuijpers}, J., {van der Post}, P., \& {Slottje}, C. 1981, \aap, 103, 331

\bibitem[{{Kumar} {et~al.}(2016){Kumar}, {Nakariakov}, \& {Cho}}]{Kumar2016}
{Kumar}, P., {Nakariakov}, V.~M., \& {Cho}, K.-S. 2016, \apj, 822, 7

\bibitem[{{Lyubchyk} \& {Voitenko}(2014)}]{Lyubchyk2014}
{Lyubchyk}, O., \& {Voitenko}, Y. 2014, Annales Geophysicae, 32, 1407

\bibitem[{{Machado} {et~al.}(1980){Machado}, {Avrett}, {Vernazza}, \&
  {Noyes}}]{Machado1980}
{Machado}, M.~E., {Avrett}, E.~H., {Vernazza}, J.~E., \& {Noyes}, R.~W. 1980,
  \apj, 242, 336

\bibitem[{{Miller} {et~al.}(1997){Miller}, {Cargill}, {Emslie}, {Holman},
  {Dennis}, {LaRosa}, {Winglee}, {Benka}, \& {Tsuneta}}]{Miller1997}
{Miller}, J.~A., {Cargill}, P.~J., {Emslie}, A.~G., {et~al.} 1997, \jgr, 102,
  14631

\bibitem[{{Moghaddam-Taaheri} \& {Goertz}(1990)}]{Moghaddam1990}
{Moghaddam-Taaheri}, E., \& {Goertz}, C.~K. 1990, \apj, 352, 361

\bibitem[{{N{\'u}{\~n}ez} {et~al.}(2005){N{\'u}{\~n}ez}, {Fidalgo}, {Baena}, \&
  {Morales}}]{Nunez2005}
{N{\'u}{\~n}ez}, M., {Fidalgo}, R., {Baena}, M., \& {Morales}, R. 2005, Annales
  Geophysicae, 23, 3129

\bibitem[{{Reale}(2010)}]{Reale2010}
{Reale}, F. 2010, Living Reviews in Solar Physics, 7, 5

\bibitem[{{Reznikova} {et~al.}(2009){Reznikova}, {Melnikov}, {Shibasaki},
  {Gorbikov}, {Pyatakov}, {Myagkova}, \& {Ji}}]{Reznikova2009}
{Reznikova}, V.~E., {Melnikov}, V.~F., {Shibasaki}, K., {et~al.} 2009, \apj,
  697, 735

\bibitem[{{Schmahl} {et~al.}(1986){Schmahl}, {Webb}, {Woodgate}, {Waggett},
  {Bentley}, {Hurford}, {Schadee}, {Schrijver}, {Harrison}, \&
  {Martens}}]{Schmahl1986}
{Schmahl}, E.~J., {Webb}, D.~F., {Woodgate}, B., {et~al.} 1986, in Energetic
  Phenomena on the Sun, ed. M.~{Kundu} \& B.~{Woodgate}

\bibitem[{{Shchukina} \& {Trujillo Bueno}(2013)}]{Shchukina2013}
{Shchukina}, N.~G., \& {Trujillo Bueno}, J. 2013, in IAU Symposium, Vol. 294,
  Solar and Astrophysical Dynamos and Magnetic Activity, ed. A.~G.
  {Kosovichev}, E.~{de Gouveia Dal Pino}, \& Y.~{Yan}, 107--118

\bibitem[{{Sirenko} {et~al.}(2000){Sirenko}, {Voitenko}, {Goossens}, \&
  {Yukhimuk}}]{Sirenko2000}
{Sirenko}, O., {Voitenko}, Y., {Goossens}, M., \& {Yukhimuk}, A. 2000, in
  American Institute of Physics Conference Series, Vol. 537, Waves in Dusty,
  Solar, and Space Plasmas, ed. F.~{Verheest}, M.~{Goossens}, M.~A. {Hellberg},
  \& R.~{Bharuthram}, 287--294

\bibitem[{{Somov}(1994)}]{Somov1994}
{Somov}, B.~V. 1994, {Fundamentals of Cosmic Electrodynamics}, Vol. 191,
  doi:10.1007/978-94-011-1184-3

\bibitem[{{Tsap} \& {Kopylova}(2017)}]{Tsap2017}
{Tsap}, Y.~T., \& {Kopylova}, Y.~G. 2017, Geomagnetism and Aeronomy, 57, 996

\bibitem[{{Tsiklauri}(2011)}]{Tsiklauri2011}
{Tsiklauri}, D. 2011, Phys. Plasmas, 18, 092903

\bibitem[{{Vernazza} {et~al.}(1981){Vernazza}, {Avrett}, \&
  {Loeser}}]{Vernazza1981}
{Vernazza}, J.~E., {Avrett}, E.~H., \& {Loeser}, R. 1981, \apjs, 45, 635

\bibitem[{{Voitenko} \& {Goossens}(2002)}]{Voitenko2002}
{Voitenko}, Y., \& {Goossens}, M. 2002, \solphys, 206, 285

\bibitem[{{Willes} \& {Robinson}(1996)}]{Willes1996}
{Willes}, A.~J., \& {Robinson}, P.~A. 1996, \apj, 467, 465

\bibitem[{{Wu} \& {Fang}(2007)}]{Wu2007}
{Wu}, D.~J., \& {Fang}, C. 2007, \apj, 659, L181

\bibitem[{{Xu} \& {Emslie}(2007)}]{Xu2007}
{Xu}, Y., \& {Emslie}, A.~G. 2007, Bulletin of the American Astronomical
  Society, 39, 213

\bibitem[{{Yukhimuk} {et~al.}(2000){Yukhimuk}, {Fedun}, {Sirenko}, \&
  {Voitenko}}]{Yukhimuk2000}
{Yukhimuk}, A., {Fedun}, V., {Sirenko}, O., \& {Voitenko}, Y. 2000, in American
  Institute of Physics Conference Series, Vol. 537, Waves in Dusty, Solar, and
  Space Plasmas, ed. F.~{Verheest}, M.~{Goossens}, M.~A. {Hellberg}, \&
  R.~{Bharuthram}, 311--316

\bibitem[{{Yukhimuk} {et~al.}(1998){Yukhimuk}, {Voitenko}, {Fedun}, \&
  {Yukhimuk}}]{Yukhimuk1998}
{Yukhimuk}, V., {Voitenko}, Y., {Fedun}, V., \& {Yukhimuk}, A. 1998, Journal of
  Plasma Physics, 60, 485

\bibitem[{{Zaitsev} \& {Stepanov}(2018)}]{Zaitsev2018}
{Zaitsev}, V.~V., \& {Stepanov}, A.~V. 2018, Geomagnetism and Aeronomy, 58, 831

\bibitem[{{Zaitsev} {et~al.}(1994){Zaitsev}, {Stepanov}, \&
  {Tsap}}]{Zaitsev1994}
{Zaitsev}, V.~V., {Stepanov}, A.~V., \& {Tsap}, Y.~T. 1994, Kinematika i Fizika
  Nebesnykh Tel, 10, 3

\bibitem[{{Zharkova} {et~al.}(2011){Zharkova}, {Kashapova}, {Chornogor}, \&
  {Andrienko}}]{Zharkova2011b}
{Zharkova}, V.~V., {Kashapova}, L.~K., {Chornogor}, S.~N., \& {Andrienko},
  O.~V. 2011, \mnras, 411, 1562

\end{thebibliography}
\bibliographystyle{aasjournal}
\end{article} 
\end{document}